\documentclass[aps,prd,showkeys,amssymb,amsfonts,
preprintnumbers,nofootinbib,twocolumn]{revtex4}
\usepackage{graphicx}
\usepackage{latexsym,url}
\usepackage{hyperref}
\usepackage{bm}
\newcommand{\be}{\begin{equation}}
\newcommand{\ee}{\end{equation}}
\newcommand{\bea}{\begin{eqnarray}}
\newcommand{\eea}{\end{eqnarray}}
\newcommand{\bref}[1]{(\ref{#1})}

\begin{document}
\begin{figure}[t]
\vspace{-1.5cm}
\hspace{-7.5cm}
\scalebox{0.085}{\includegraphics{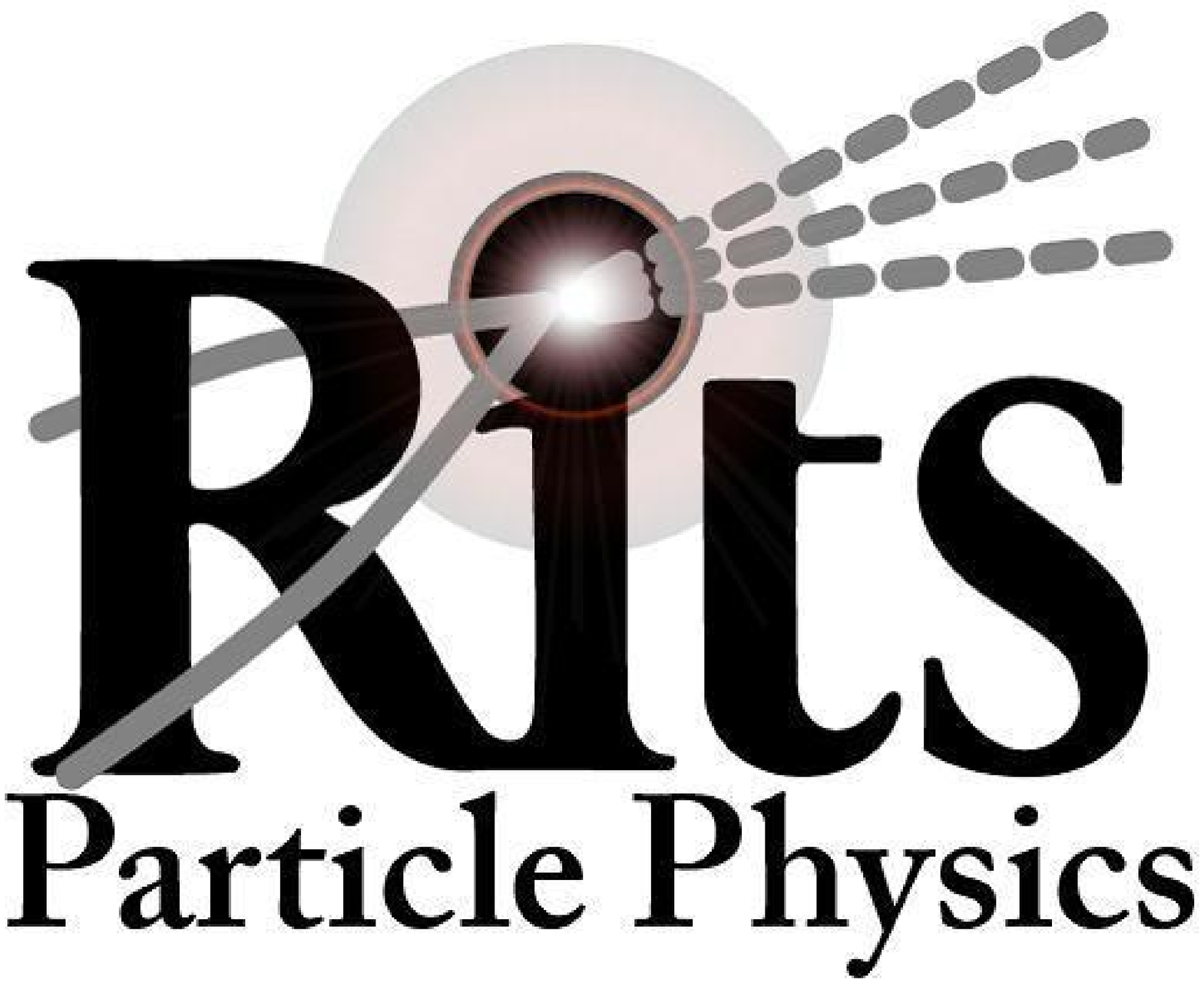}} 
\end{figure}
\title{Electroweak inflation and reheating in the NMSSM}
\author{Takeshi~Fukuyama}
\email[E-mail: ]{fukuyama@se.ritsumei.ac.jp}
\affiliation{Department of Physics, Ritsumeikan University, 
Kusatsu, Shiga 525-8577, Japan}
\author{Tatsuru~Kikuchi}\email[E-mail: ]{rp009979@se.ritsumei.ac.jp}
\affiliation{Department of Physics, Ritsumeikan University, 
Kusatsu, Shiga 525-8577, Japan}
\author{Wade~Naylor}
\email[E-mail: ]{naylor@se.ritsumei.ac.jp}
\affiliation{Department of Physics, Ritsumeikan University, 
Kusatsu, Shiga 525-8577, Japan}
\begin{abstract}
A low reheating temperature is now well motivated by the recently 
reconsidered gravitino problem, if we incorporate supergravity. 
In this article, we propose a model which naturally realizes a low 
reheating temperature. The model is based on the next to minimal 
supersymmetric standard model (NMSSM) by identifying a singlet scalar 
field as the inflaton. This entertains the possibility that the inflaton 
may be detected at future colliders, such as the LHC.
\end{abstract}
%\pacs{04.50.+h; 98.80.Cq}
\preprint{RITS-PP-008}
\date{\today}
\maketitle
\section{Introduction}
The supersymmetric (SUSY) extension of the standard model (the MSSM) 
is one of the most promising ways to solve the gauge hierarchy problem 
in the standard model \cite{SUSY}. Though introducing supersymmetry into 
the Standard Model solves a lot of problems which could not be solved 
in the Standard Model, some still remain some. 
In confrontation with cosmology we have unnecessary thermal relics: 
the {\it moduli}, {\it gravitino}, and so on. These would ruin a successful 
prediction of the photon number density at the recombination era 
in the standard Big Bang cosmology. 

However, if thermalization occurs at a sufficiently low scale we have 
no such unnecessary thermal relics. Hence, we are driven to construct 
a model with a low reheat temperature. Though there have been some studies 
on low reheat temperature models \cite{low} the reheat temperature was just 
put in by hand. However, in this letter we shall attempt to realize it by using 
a non-perturbative mechanism, instant preheating \cite{instant}.

On the other hand, there is another problem in the MSSM: 
the so called $\mu$ problem, arising from SUSY preserving mass terms 
for the Higgs fields in the superpotential, $\mu H_u H_d$ \cite{mu}. 
If the MSSM is viable at the (reduced) Planck scale, 
$M_P \simeq 2.4 \times 10^{18}$ GeV, the $\mu$ parameter is naturally 
expected to lie at the Planck scale. 
However, the $\mu$ parameter should be at the electroweak scale 
to provide the correct mass scale for the gauge bosons. This may be solved by adding an additional singlet to the MSSM: the next to 
minimal supersymmetric standard model (NMSSM) \cite{NMSSM, NMSSM2}. 

In this letter, we shall discuss inflation in the NMSSM, especially with regard to realising a low reheating temperature. 
For some related work in the $\phi$NMSSM model, see \cite{King}.
%%%%%%%%%%%%%%%%%%%%%%%%%%%%%%%%%%
\section{The NMSSM}
This model uses the following set of Higgs fields
having the indicated $SU(2)_L \times U(1)_Y$ charges,
\bea
H_d &=& \left(
\begin{array}{c}
H_d^0 \\
H_d^- 
\end{array}
\right) \sim ({\bf 2},1),~
H_u =
\left(
\begin{array}{c}
H_u^+ \\
H_u^0
\end{array}
\right) \sim ({\bf 2},-1),\nonumber\\
S &\sim& ({\bf 1},0 )
\eea
and we use the following superpotential for the Higgs fields
to describe the effective theory below the Planck scale
\be
W =\lambda S H_u \cdot H_d + \frac{1}{3}\kappa S^3\;,
\label{W}
\ee
where the product of the $SU(2)$ doublets are defined by, e.g.
\be
H_u \cdot H_d \equiv H_u^+ H_d^- - H_u^0 H_d^0 \;.
\ee
For the soft SUSY breaking terms for the Higgs fields we take
\bea
-{\cal L}_{\mbox{soft}} &=&
m_{H_u}^2 |H_u|^2 + m_{H_d}^2 |H_d|^2 + m_{S}^2 |S|^2
\nonumber\\
&+& \lambda A_\lambda S H_u \cdot H_d
+\frac{1}{3} \kappa A_\kappa S^3 +h.c.
\eea
It is a well known fact that the superpotential (\ref{W}) has a discrete 
${\mathbb Z}_3$ symmetry, which would be at odds with cosmology by 
producing unwanted topological defects if this ${\mathbb Z}_3$ symmetry 
is local and spontaneously broken. However, here we shall assume 
it just a global symmetry.

After developing a vacuum expectation value (VEV) for $S$,
we obtain a natural explanation for the electroweak scale $\mu$-term,
\be
\mu_{\rm{eff}} \equiv \lambda \left<S\right> \sim M_Z\;.
\ee
From the Higgs part of the superpotential \bref{W} and taking the $D$-term 
contributions into account, we obtain the following Higgs potential,
\newpage
\bea
V &=& \lambda^2 \left(|H_u|^2 |S|^2 + |H_d|^2 |S|^2 
+ |H_u \cdot H_d|^2 \right) 
\nonumber\\
&+& \kappa^2 |S^2|^2
+ \lambda \kappa \left((S^*)^2 H_u \cdot H_d  + h.c. \right)
\nonumber\\
&+& \frac{g^2 +{g^\prime}^2}{8} \left(|H_u|^2 - |H_d|^2 \right)^2
+ \frac{g^2}{2} |H_u^{+} H_d^{0*} + H_u^{0} H_d^{-*} |^2 
\nonumber\\
&+& m_{H_u}^2 |H_u|^2 + m_{H_d}^2 |H_d|^2 + m_{S}^2 |S|^2
\nonumber\\
&+& \lambda A_\lambda S H_u \cdot H_d 
+ \frac{1}{3} \kappa A_\kappa S^3 +h.c.
\eea
The first and second line are the F-terms, the third and fourth line corresponds to the D-terms and the last line contains the soft SUSY breaking terms. To obtain the vacuum condition, we take the Higgs fields at the potential minimum to be
\be
\left<H_d \right> = \left(
\begin{array}{c}
v \cos \beta \\
0
\end{array}
\right) \;,~~
\left<H_u \right> =
\left(
\begin{array}{c}
0 \\
v \sin \beta
\end{array}
\right) \;,~~
\left<S\right> = s \;,
\ee
where $v=174$~[GeV].

From the this vacuum condition, we obtain three relations, 
linking the three soft mass parameters to the three VEV's of 
the Higgs fields.
\bea
m_{H_u}^2
&=&
-\frac{g^2 +{g^\prime}^2}{4} v^2 \cos 2 \beta
-\lambda^2 v^2 \sin^2 \beta 
\nonumber\\
&+& \left(A_\lambda + \kappa s \right) \lambda s \tan \beta
- \lambda^2 s^2 \;,
\label{v1}
\nonumber\\
m_{H_d}^2
&=&
+\frac{g^2 +{g^\prime}^2}{4} v^2 \cos 2 \beta
-\lambda^2 v^2 \cos^2 \beta 
\nonumber\\
&+& \left(A_\lambda + \kappa s \right) \lambda s \cot \beta
- \lambda^2 s^2 \;,
\label{v2}
\nonumber\\
m_S^2
&=&-2 \kappa^2 s^2 
- \lambda^2 v^2
+ \kappa \lambda v^2 \sin 2 \beta
\nonumber\\
&+& \lambda A_\lambda \frac{v^2 \sin 2 \beta}{s}
- \kappa A_\kappa s \;.
\label{v3}
\eea
The VEV of the singlet $S$ in the exact SUSY limit can easily found by setting $m_S=0$ with $A_\kappa=A_\lambda=0$:
\be
s^2 = \frac{\lambda v^2}{2 \kappa^2} \left(- \lambda
+ \kappa \sin 2 \beta \right) 
\ee
and hence,
\be
\mu_{\rm{eff}} = \frac{\lambda^{3/2} v}{\sqrt{2} \kappa} 
\left(- \lambda+ \kappa \sin 2 \beta \right)^{1/2} \sim v \;.
\ee
So, if we have both parameters $\lambda$ and $\kappa$ of order one,
$\lambda \sim \kappa \sim 1$, we can find a solution to the $\mu$ problem.

The mass matrices of the Higgs fields can be read off from
the above potential by expanding the Higgs fields around their minima.
For the charged Higgs fields, we have
\bea
H_d^- &=& H^- \sin \beta - G^- \cos \beta \;,
\nonumber\\
H_u^+ &=& H^+ \cos \beta + G^+ \sin \beta \;,
\eea
where $G^{\pm}$ become would-be Goldstone bosons.
For the neutral Higgs fields, we have
\bea
\Im(H_d^0) &=& \frac{1}{\sqrt{2}} \left(P_1 \sin \beta 
- G^0 \cos \beta \right) \;,
\nonumber\\
\Im(H_u^0) &=& \frac{1}{\sqrt{2}} \left(P_1 \cos \beta 
+ G^0 \sin \beta \right) \;.
\nonumber\\
\Im(S) &=& \frac{P_2}{\sqrt{2}} \;,
\eea
where $G^0$ becomes a would-be Goldstone boson, and
\bea
\Re(H_d^0) &=& v \cos \beta + \frac{1}{\sqrt{2}}
\left(-S_1 \sin \beta + S_2 \cos \beta \right) \;,
\nonumber\\
\Re(H_u^0) &=& v \sin \beta + \frac{1}{\sqrt{2}}
\left(S_1 \cos \beta + S_2 \sin \beta \right) \;.
\nonumber\\
\Re(S) &=& s + \frac{\sigma}{\sqrt{2}} \;.
\eea
Then, the charged Higgs mass spectra can explicitly be written 
since they are already in their mass eigenstates.
\bea
M_{H^{\pm}}^2 &=& M_A^2 + M_Z^2 - \frac{1}{2} (\lambda v)^2 \;,
\nonumber\\
M_A^2 &=& \frac{\lambda s}{\sin 2 \beta} 
\left(\kappa s + \sqrt{2} A_\lambda \right)\;.
\eea
The neutral Higgs fields $P_1,~P_2$ and $S_i~(i=1,2,3)$ are not
in their mass eigenstates. Their mass matrices are given as follows.
For the Higgs fields $P_1,~P_2$,
\bea
({\cal M}_{P}^2)_{11} &=& M_A^2 \;,
\nonumber\\
({\cal M}_{P}^2)_{12} &=& \frac{1}{2}\left(M_A^2 \sin 2\beta 
- 3 \lambda \kappa s^2 \right) \left(\frac{v}{s} \right) \;,
\nonumber\\
({\cal M}_{P}^2)_{22} &=& \frac{1}{4}\left(M_A^2 \sin 2\beta 
+ 3 \lambda \kappa s^2 \right) \left(\frac{v}{s}\right)^2 \sin 2\beta 
\nonumber\\
&-& \frac{3}{\sqrt{2}} \kappa s A_\kappa \;,
\eea
and for the Higgs fields $S_i~(i=1,2,3)$,
\bea
({\cal M}_{S}^2)_{11} &=& M_A^2 + 
\left\{M_Z^2 - \frac{1}{2} (\lambda v)^2 \right\} \sin^2 2\beta \;,
\nonumber\\
({\cal M}_{S}^2)_{12} &=& -\frac{1}{2}
\left\{M_Z^2 - \frac{1}{2} (\lambda v)^2 \right\} \sin 4\beta \;,
\nonumber\\
({\cal M}_{S}^2)_{13} &=& -\frac{1}{2}
\left(M_A^2 \sin 2\beta + \lambda \kappa s^2 \right) 
\left(\frac{v}{s}\right) \cos 2\beta \;,
\nonumber\\
({\cal M}_{S}^2)_{22} &=& M_Z^2 \cos^2 2\beta 
+ \frac{1}{2} (\lambda v)^2 \sin^2 2\beta \;,
\nonumber\\
({\cal M}_{S}^2)_{23} &=& \frac{1}{2}
\left(2 \lambda^2 s^2 - M_A^2 \sin^2 2\beta 
- \lambda \kappa s^2 \sin 2\beta \right) \left(\frac{v}{s} \right) \;,
\nonumber\\
({\cal M}_{S}^2)_{33} &=& \frac{1}{4} M_A^2 \sin^2 2\beta 
\left(\frac{v}{s} \right)^2 + 2 \kappa^2 s^2 
\nonumber\\
&+& \frac{1}{\sqrt{2}} \kappa s A_\kappa 
- \frac{1}{4} \lambda \kappa v^2 \sin 2\beta \;.
\eea
From here, we see that the lightest (neutral) Higgs boson mass is
bounded by
\be
m_{h}^2 \lesssim M_Z^2 \cos^2 2\beta +
(\lambda v)^2 \sin^2 2\beta \;,
\ee
which relaxes the tree level MSSM upper bound, 
$m_{h}^2 \lesssim M_Z^2 \cos^2 2\beta$.
For a detailed calculation tool for all the Higgs mass spectra
in the NMSSM open source code can be found in \cite{NMHDECAY}.

%%%%%%%%%%%%%%%%%%%%%%%%%%%%
\section{Inflation at the Electroweak Scale}
We shall now consider how to construct an electroweak scale inflation model, 
which can avoid the gravitino problem in local SUSY models \cite{weinberg}. 
For previous works on such model building, see 
\cite{Knox:1992iy, Kinney:1995cc, Garcia-Bellido:1999sv, Krauss:1999ng, Lyth, 
German:2001tz, mix}.

Our model of low scale inflation is based on the NMSSM, and hereafter, 
we regard $S$, which provides a natural explanation for the $\mu$ term, 
as the inflaton field. One major reason being that ``inflaton-Higgs mixing'' 
\cite{mix} occurs in this model and thus, the inflaton itself might 
be detectable at a collider, e.g. the LHC.

\begin{figure}[t]
\begin{center}
\scalebox{0.95}{\includegraphics{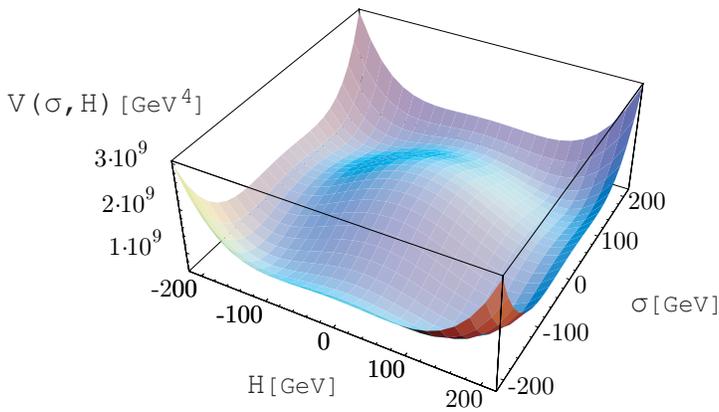}}
\caption{The potential with a parameter set:
$s=174~{\rm GeV},~\lambda = \kappa =1$. The other parameters
are suitably chosen so as to satisfy the vacuum condition.}
\label{Fig1}
\end{center}
\end{figure}

\begin{figure}[t]
\begin{center}
\scalebox{0.8}{\includegraphics{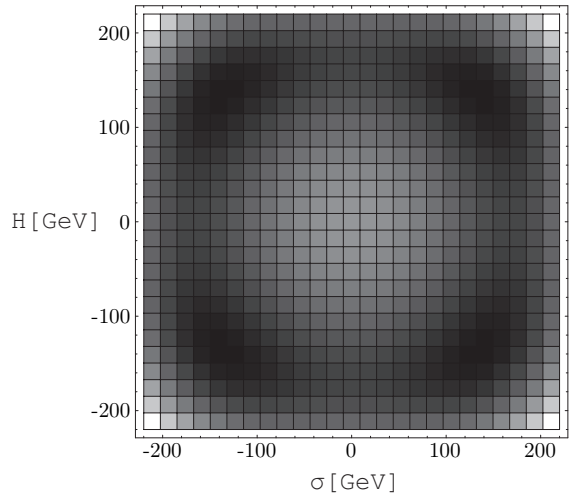}} 
\caption{A view of the potential from above in Fig.~\ref{Fig1}. 
Here the height of the potential is illustrated by the density of the plot. 
The four dark regions are the minimum points.}
\label{Fig2}
\end{center}
\end{figure}

Let us remind the reader that the scalar potential for the inflaton field,
$\sigma \equiv \sqrt{2}\left[\Re (S) -s \right]$, and for the Higgs fields,
$h_u \equiv \sqrt{2}\left[\Re(H_u^0) -v \sin\beta \right]$,
$h_d \equiv \sqrt{2}\left[\Re(H_d^0) -v \cos\beta \right]$
has the following form:
\bea
V(\sigma,h_u,h_d)
&=& \frac{1}{2}
\left[-\lambda (-\lambda + \kappa \sin 2\beta)v^2 
+ 6 \kappa^2 s^2 + m_{S}^2 
\right. \nonumber\\
&+&\left. \kappa A_\kappa s \right] \sigma^2 
+ \frac{\kappa^2}{4} \sigma^4
\nonumber\\
&+& \frac{\lambda^2}{4} \left(h_u^2 + h_d^2 \right) \sigma^2
- \frac{\lambda \kappa}{2} h_u h_d \sigma^2
\nonumber\\
&+& \frac{\lambda^2}{4} \left(h_u h_d \right)^2
+ \frac{g^2 +{g^\prime}^2}{32} \left(h_u^2 - h_d^2 \right)^2
\nonumber\\
&+& \frac{1}{2} \left(m_{H_u}^2 + \lambda^2 s^2 \right) h_u^2
\nonumber\\
&+& \frac{1}{2} \left(m_{H_d}^2 + \lambda^2 s^2 \right) h_d^2
\nonumber\\
&-& \lambda s \left(\kappa s + A_\lambda \right) h_u h_d \;.
\label{V}
\eea
In Fig. \ref{Fig1} we illustrate the potential as a function of $H$ and $\sigma$, with the view from above given in Fig. \ref{Fig2}. %This {\it hybrid} like potential has four degenerate minima, but we shall assume that the inflaton field rolls from the origin into only one of the minima. 

%%%%%%%%%%%%%%%%%%%%%%%%%%%%%%%%%%%%%%%%%%%%%%%
%%%%%%%%%%%%%%%%%%%%%%%%%%%%%%%%%%%%%%%%%%%%%%%
%%%%%%%%%%%%%%%%%%%%%%%%%%%%%%%%%%%%%%%%%%%%%%%

From the potential (\ref{V}) we can calculate the necessary conditions 
for the (hybrid) inflation model to become viable \cite{Lyth}. 
First of all, there are two necessary conditions for inflation, 
the so called flatness and the slow-roll conditions. 
These can be illustrated by defining
\be
\epsilon \equiv \frac{M_P^2}{2} 
\left(\frac{\left<V' \right>}{\left<V \right>}\right)^2 \;,
~~
\eta \equiv M_P^2 \left|\frac{\left<V'' \right>}{\left<V \right>} \right| \;,
\ee
where $'$ represents the derivative with respect to the inflaton field,
$\sigma$. Clearly, the necessary conditions are
\be
\epsilon \ll 1 \;,~~
\eta \ll 1 \;.
\ee
To evaluate the parameter $\eta$, we need to calculate $V''$ 
from the potential given by (\ref{V}):
\be
\left<V'' \right> = \left[- \lambda (-\lambda + \kappa \sin 2\beta)v^2
+ 6 \kappa^2 s^2 + m_{S}^2 + \kappa A_\kappa s \right] \;.
\ee
Then the required condition for the $\eta$ parameter becomes
\be
\eta = \frac{M_P^2}{\left<V \right>} 
\left[- \lambda (-\lambda + \kappa \sin 2\beta)v^2
+ 6 \kappa^2 s^2 + m_{S}^2 + \kappa A_\kappa s \right] \;.
\ee
To keep an inflationary era for long enough, the number of e-foldings 
for the inflaton field should be $N > 60$. The field value $\sigma_N$ 
with $N$ e-foldings is given by
\be
N=\frac{1}{M_P^2} \int_{\sigma_{\rm end}}^{\sigma_N} d\sigma 
\frac{V}{V^\prime} \;,
\ee
where $\sigma_{\rm end}$ is the value of the field at which 
the inflation ends. In hybrid inflationary models, it is also useful 
to define the parameter
\be
\eta_{H} \equiv M_P^2 \left(\frac{M_H^2}{\left<V \right>}\right) 
\simeq 10^{32} \;.
\ee
Then, the CMB anisotropy in low scale inflation models
requires the coupling constant, $\lambda$, to satisfy \cite{Lyth}
\bea
\lambda^2 &=& 3 \times 10^{-7} \eta^2 \eta_H
\nonumber\\
&\simeq& 3 \times 10^{25} \eta^2 \;.
\eea
In order to realize instant preheating effectively (see the next section), we need a marginal coupling constant $\lambda \sim 1$, hence the parameter
$\eta$ should be $\eta \simeq 10^{-25}$. This requires the potential
to be very flat in the inflaton field direction:
\be
\left<V'' \right> \simeq 10^{-40} ~{\rm GeV^2} \:.
\ee
In the first order approximation this reduces to the following condition
\be
- \lambda (-\lambda + \kappa \sin 2\beta)v^2
+ 6 \kappa^2 s^2 + m_{S}^2 + \kappa A_\kappa s = 0 \;.
\ee
Putting the vacuum condition (\ref{v3}) into this equation gives
a very simple form:
\be
4 \kappa^2 s^2 + \lambda A_\lambda \frac{v^2 \sin 2 \beta}{s} = 0 \;.
\ee
Here we assume the parameter $\kappa$ to be of order one, then the above 
equation determines the parameter $A_\lambda$ as
\be
A_\lambda = \frac{- 4 \kappa^2 s^3}{\lambda v^2 \sin 2 \beta} 
\sim - \frac{4 v}{\sin 2 \beta} \;.
\ee
In such a case we obtain electroweak inflation which explains 
the density perturbations and also gives a viable mechanism 
for instant preheating, which is discussed in the next section.

%%%%%%%%%%%%%%%%%%%%%%%%%%%%%%%%%%%%%
\section{Instant Preheating}
The instant preheating mechanism \cite{instant} requires the particle 
decay channel $\sigma \to H \to f \bar{f}$, where $H$ is one of the CP-even 
Higgs bosons, which can have a mass much heavier than $\sigma$ due to 
preheating, and $f$ denotes a fermion with mass lighter than $H$. This
leads to a very natural explanation for a low reheating temperature
\be
T_R \simeq 0.05 M_H \lesssim 50~{\rm [GeV]} \;.
\ee
Importantly, this also satisfies the gravitino constraint on the reheating 
temperature \cite{gravitino},
\be
T_R \lesssim 100~{\rm [TeV]}
\ee
for the TeV scale gravitino $m_{3/2} \simeq 1$ [TeV].

The key ingredient for obtaining the relation $T_R \sim M_H$ was given in
\cite{FKN}. Given an inflationary model, we can investigate the effects 
of preheating to generate a large decay rate for the inflaton, which can be 
achieved by using the instant preheating mechanism \cite{instant}. 
In this case the inflaton oscillates about the minimum of the potential 
only once and it is possible to show that \cite{instant,pre}
\be
n_k = \exp \left(-{\pi (k^2/a^2+ M_{H}^2) \over \lambda s M_{H}}
 \right)~.
\ee
As discussed in \cite{instant} $s M_{H}$ can be replaced by 
$|\dot{\sigma}(t)|$, which leads to
\be
n_k = \exp\left(-{\pi (k^2/a^2+ M_{H}^2) \over \lambda |\dot{\sigma}(t)|}
\right) \;.
\ee
This can then be integrated to give the number density for the Higgs field,
\bea
\label{suppr}
n_{H} &=& {1 \over 2\pi^2} \int \limits_0^{\infty} dk \,k^2 n_k
\ =\ {({\lambda \dot{\sigma}(t)})^{3/2} \over 8\pi^3}~
\exp \left(-{\pi  M_{H}^2  \over  \lambda |\dot{\sigma}(t)|}\right)
\nonumber\\
&=&
{\left(\lambda s M_{H} \right)^{3/2} \over 8\pi^3}~
\exp \left(-{\pi  M_{H} \over \lambda s}\right)
\ \simeq\ {M_{H}^3 \over 8\pi^3}\,e^{-\pi} \,.
\eea
As argued in \cite{instant}, if the couplings are of order 
$\lambda \sim 1$ then there need not be an exponential suppression of 
the number density. Interestingly, this fact has recently been used in a
model of non-thermal leptogenesis \cite{Ahn}. Whence, the resultant reheating temperature is found to be
\bea
T_R &=& 
\left(\frac{30}{g_* \pi^2} \cdot 
M_{H} \cdot n_{H} \right)^{1/4}
\ \simeq\
\left(\frac{15}{4 \pi^5 g_*} \right)^{1/4} M_{H} e^{-\pi/4}
\nonumber\\
&\cong& 0.05 \times M_{H} \;.
\label{pretemp}
\eea
It should be stressed that the reheating temperature above, obtained from the preheating mechanism, is proportional 
to the mass of the decayed particle, the Higgs boson mass 
($T_R \propto M_H$) and does not depend on the inflaton mass.

%%%%%%%%%%%%%%%%%%%%%%%%%%%%%%%%%%%%%
\section{Discussion}
Although there are various problems in the MSSM, we have shown that 
the simplest extension of it, the NMSSM, has interesting consequences 
for cosmology. In particular, by assuming that the extra singlet in 
the NMSSM is the inflaton field we can {\it naturally} realize inflation 
and a low reheat temperature which is not in conflict with the gravitino relic abundance bound.

As far as electroweak baryogenesis is concerned, even in such a low reheat temperature model, the NMSSM should naturally incorporate a strongly first order phase transition with less stringent bounds than in the MSSM \cite{Trodden}.

Moreover, although we require a fine tuning in the slow-roll conditions, such a fine tuning {\it only} affects the soft SUSY breaking terms in the Lagrangian, which are themselves put in by hand. This feature might be considered as a relaxation of the fine tuning problem, that is, we are only tuning the soft SUSY breaking terms. 

Finally, given that we are considering an NMSSM model, where the scalar-singlet is identified as the inflaton, we might be lead to believe that the inflaton could be detected at future colliders, such as the LHC.

%%%%%%%%%%%%%%%%%%%%%%%%%%%%%%%%%%%%%%%%%
\section*{Acknowledgments}
The work of TF is supported in part by 
the Grant-in-Aid for Scientific Research from the Ministry 
of Education, Science and Culture of Japan (\#16540269).
The work of TK is supported by the Research Fellowship 
of the Japan Society for the Promotion of Science (\#7336).

%%%%%%%%%%%%%%%%%%%%%%%%%%%%%%%%%%%%%%%%%%%

\end{document}